\documentclass[aps,nofootinbib,superscriptaddress]{revtex4}

\usepackage{graphicx}
\usepackage{bm}
\usepackage{amsmath}
\usepackage{amssymb}
\usepackage{hyperref}

\newcommand{\qslash}{\kern 0.2 em n\kern -0.50em /}
\newcommand{\nslash}{\kern 0.2 em n\kern -0.50em /}
\newcommand{\kslash}{\kern 0.2 em k\kern -0.45em /}
\newcommand{\lslash}{\kern 0.2 em l\kern -0.50em /}
\newcommand{\pslash}{\kern 0.2 em p\kern -0.50em /}
\newcommand{\Sslash}{\kern 0.2 em S\kern -0.50em /}
\newcommand{\Pslash}{\kern 0.2 em P\kern -0.50em /}
\newcommand{\Dslash}{\kern 0.2 em D\kern -0.65em /\kern 0.15em}

\newcommand{\ii}{i}

\usepackage{color}

\begin{document}

\title{Double Collins effect in $e^+ e^-\to\Lambda \bar\Lambda X$ process in a diquark spectator model}
\author{Xiaoyu Wang}
\affiliation{School of Physics, Southeast University, Nanjing 211189, China}
\author{Yongliang Yang}\affiliation{School of Physics, Southeast University, Nanjing
211189, China}
\author{Zhun Lu}\email{zhunlu@seu.edu.cn}\affiliation{School of Physics, Southeast University, Nanjing
211189, China}

\begin{abstract}
We study the Collins function $H^\perp_{1}$ of the $\Lambda$ hyperon, which describes the fragmentation of a transversely polarized quark into an unpolarized $\Lambda$ hyperon. We calculate $H^\perp_{1}$ for light quarks of the $\Lambda$ hyperon, in the diquark spectator model with a Gaussian form factor for the hyperon-quark-diquark vertex. The model calculation includes contributions from both the scalar diquark and vector diquark spectators. Using the model result, we estimate the weighted $\cos2\phi_0$ asymmetry in the process $e^+e^- \to \Lambda\bar{\Lambda}X$ contributed by the coupling of two Collins functions. The QCD evolution effects for the first $k_T$-moment of the Collins function and the unpolarized fragmentation function $D_1(z)$ are also included. The results show that asymmetry is sizable and measurable at the kinematical configurations of Belle and BaBar experiments. We also find that the evolution effects play an important role in the phenomenological analysis.
\end{abstract}


\maketitle

\section{introduction}

The Collins function~\cite{Collins:1992kk} $H_1^\perp(z,k_T^2)$ is a novel transverse momentum dependent (TMD) fragmentation function which encodes the correlation between the transverse spin of the fragmenting quark and the transverse momentum of the unpolarized final-state hadron.
As a Time-reversal-odd (T-odd) function, Collins function can be served as a quark spin analyzer and also be used to explore the non-perturbative fragmentation mechanism of hadrons.
The experimental measurements of the pion Collins function came from several single transverse spin asymmetries in semi-inclusive deep inelastic scattering~(SIDIS)~\cite{Airapetian:2004tw,Airapetian:2010ds,Adolph:2012sn,Adolph:2014zba,
Ageev:2006da,Alekseev:2008aa,Alekseev:2010rw} from the HERMES and the COMPASS Collaborations, and the azimuthal asymmetry in $e^+e^-$ annihilating process~\cite{Abe:2005zx,Seidl:2008xc,TheBABAR:2013yha,Ablikim:2015pta,Efremov:1998vd} from the BaBar and Belle Collaborations.
Combining the experimental data from SIDIS and $e^+e^-$ annihilating processes, one can extract the Collins function as well as the transversity distribution function~\cite{Anselmino:2007fs,Anselmino:2008jk,Anselmino:2013vqa,Anselmino:2015sxa}, which makes the Collins function a useful tool to investigate the internal structure for hadrons.
Recently, the azimuthal asymmetry of charged kaon pair production in $e^+ e^-$ annihilation was measured by the BaBar Collaboration~\cite{Aubert:2015hha}, making the extraction~\cite{Anselmino:2015fty} of the kaon Collins function possible.
In addition, several model calculations of the Collins functions of the pion and kaon have been presented in Refs.~\cite{Bacchetta:2007wc,Amrath:2005gv,Gamberg:2003eg,Bacchetta:2003xn,Bacchetta:2002tk,Bacchetta:2001di} and used to make predictions on the physical observables~\cite{Schweitzer:2003yr,Gamberg:2003pz,Gamberg:2003eg}.

Although in the past a lot of experimental data and theoretical analyses have provided information about the Collins functions for pion and kaon mesons, knowledge about the Collins function of the $\Lambda$ hyperon is much more limited.
Meanwhile there are increasing interests on the novel fragmentation mechanism of the $\Lambda$ hyperon, as it is partly responsible to the observed spin polarization or spin transfer of the spin-$1/2$ Lambda hyperon produced in high-energy inclusive process~\cite{Heller:1978ty,Heller:1983ia,Ramberg:1994tk,Smith:1986fz,Lundberg:1989hw,Pondrom:1985aw,
Fanti:1998px,Agakishiev:2014kdy,ATLAS:2014ona}.
A T-odd spin-dependent TMD fragmentation $D_{1T}^\perp(z,k_T^2)$, which describes the number density of a transversely polarized $\Lambda$ hyperon fragmented from an unpolarized quark, is found to play an important role in this aspect and has been studied intensively~\cite{Boer:2007nh,Anselmino:2001js,Anselmino:2000vs,Collins:1992kk,Dong:2004qs,Sivers:1989cc,Felix:1999tf}.
As the chiral-odd partner of the fragmentation function $D_{1T}^\perp(z,k_T^2)$, the Collins function of the $\Lambda$ hyperon also contains complimentary information of the lambda fragmentation and can give rise to the azimuthal asymmetries in high energy process.
In order to understand the underlying mechanism of transversely polarized quark fragmenting to unpolarized lambda, we resort to model calculation to acquire the knowledge of the corresponding non-perturbative quantity, which is the main goal of this work.
For this purpose, for the first time, we calculate the lambda Collins function for the up, down and strange quarks, using a spectator model~\cite{Nzar:1995wb,Jakob:1997wg}.
The spectator model has been applied to calculate the Collins functions of the pion and kaon mesons~\cite{Bacchetta:2007wc}, as well as as well as the twist-3 collinear fragmentation function of the pion~\cite{Lu:2015wja,Yang:2016mxl}, with a pseudoscalar pion-quark coupling and Gaussian form factors at the pion-quark/antiquark vertex.
In these cases the quark or antiquark is taken as the spectator system.
The calculation presented in Ref.~\cite{Bacchetta:2007wc} showed that the model resulting pion Collins function is in reasonable agreement with the available parametrization~\cite{Kretzer:2000yf}.
Recently, the spectator model has also been extended to calculate the fragmentation function $D^\perp_{1T}$ of the $\Lambda$ hyperon in Ref.~\cite{Yang:2017cwi}.
In this case the spectator system is a diquark, and the calculation includes contributions from both the scalar diquark and vector diquark.

The Collins function can enter the description in SIDIS, $e^+ e^-$ annihilation and inclusive hadron production in hadron collision.
To test the feasibility of measuring the lambda Collins function in experiments, we will study the unpolarized $e^+ e^- \to \Lambda \bar{\Lambda} X $ process, in which only fragmentation functions are involved.
In this process, the convolution of two Collins functions can generate at leading order (in the expansion of $1/Q$) a $\cos 2\phi_0$ azimuthal asymmetry~\cite{Boer:1997mf,Boer:2008fr}.
The theoretical approach to analyze the asymmetry is the TMD factorization in which the evolution formalism of the TMD function is very complicate.
To avoid the complicity, in this work we study the $q_T^2$ weighted azimuthal asymmetry in $e^+ e^- \to \Lambda \bar{\Lambda} X $, with $q_T$ the transverse momentum of the virtual photon.
In the weighted procedure the differential cross-section can be expressed as the product of the first $k_T$-moments of the Collins function $H_1^{\perp(1)}$ instead of the convolution of TMD fragmentation functions.
The corresponding experiments may be accessible in the Belle and BaBar $e^+ e^-$ facilities.
We also take into account the QCD evolution effect of $H_1^{\perp(1)}$ as the energy scale for those experiments is much larger than the model scale.

The remaining content of this paper is organized as follows.
In Sec.~\ref{sec:model}, we calculate the T-odd Collins function $H_1^{\perp}$ in the diquark spectator model by including both the scalar diquark and vector diquark spectators.
The QCD evolution effect of the first $k_T$-moment of Collins function $H_1^{\perp(1)}(z)$ is also studied.
In Sec.~\ref{Sec:weighted}, we numerically estimate the $q_T^2$ weighted Collins asymmetry at the energy scale around the Belle and BaBar kinematics by considering the QCD evolution effects of both $H_1^{\perp(1)}(z)$ and $D_1(z)$.
We summarize this work in Sec.~\ref{sec:conclusion}.

\section{Model calculation of the Collins function for $\Lambda$ hyperon}
\label{sec:model}

In this section, we calculate the Collins function $H^{\perp}_{1}(x,\bm{k}^2_T)$, which describes the number density of an unpolarized $\Lambda$ hyperon fragmented from a transversely polarized quark~\cite{Bacchetta:2004jz}:
\begin{align}
D_{\Lambda/q^\uparrow}(z,\bm{P}_{\Lambda\,T}) -  D_{\Lambda/q^\uparrow}(z,-\bm{P}_{\Lambda\,T})=
\Delta D_{\Lambda/q^\uparrow}(z,\bm{P}_{\Lambda\,T}^2){(\hat{\bm{k}} \times \bm{P}_{\Lambda\,T})\cdot \bm{S}_q \over z M_\Lambda}\,,
\end{align}
where $\bm{P}_{\Lambda\,T}$ is the transverse momentum of the $\Lambda$ hyperon with respect to the quark momentum $\bm k$, ${\bm S}_q$ is the spin vector of the fragmenting quark, and $z$ and $M_\Lambda$ are the light-cone momentum fraction and the mass of the produced $\Lambda$ hyperon, respectively .
Either $H^\perp_{1}$ or $\Delta D_{\Lambda/q^\uparrow}$ may be referred to as the Collins function defined in Refs.~\cite{Barone:2001sp,Bacchetta:2004jz,Anselmino:2000mb}.
The relation between them is
\begin{align}
\Delta D_{\Lambda/q^\uparrow}(z,\bm{k}_T^2)={2|\bm{P}_{\Lambda\,T}|\over zM_\Lambda}H^{\perp\,q}_{1}(z,\bm{k}_T^2)={2|\bm{k}_{T}|\over M_\Lambda}H^{\perp\,q}_{1}(z,\bm{k}_T^2)\,,
\end{align}
where $\bm k_T$ is related to $\bm P_{\Lambda\,T}$ by  $\bm k_T = -\bm P_{\Lambda\,T}/z$.

The Collins function can be calculated from the following trace
\begin{align}
{\epsilon_T^{\alpha\rho} k_{T\rho}\over M_\Lambda}H^\perp_{1} &={1\over 4}\textrm{Tr}[(\Delta(z,k_T;S_\Lambda) +\Delta(z,k_T;-S_\Lambda)) \ii\sigma^{\alpha-}\gamma_5] \,.
\end{align}
Here, the quark-quark fragmentation correlation function $\Delta(z,k_T;S_\Lambda)$ is defined as~\cite{Bacchetta:2008af,Bacchetta:2006tn}
 \begin{align}\label{eq:delta1}
\Delta(z,k_T;S_\Lambda)&={1\over 2z}\int{dk^+\Delta(k,P_\Lambda;S_\Lambda)}\notag\\
&\equiv \sum_X\int{d\xi^+d^2\bm{\xi_T}\over 2z(2\pi)^3} e^{ik\cdot\,\xi}\langle 0|\, {\cal U}^{n^+}_{(+{\infty},\xi)}
\,\psi(\xi)|P_\Lambda,S_\Lambda; X\rangle\langle P_\Lambda,S_\Lambda; X|\bar{\psi}(0)\,
{\cal U}^{n^+}_{(0,+{\infty})}
|0\rangle \bigg|_{\xi^-=0}\,,
\end{align}
with $k^-={P^-_\Lambda \over z}$.
The Wilson line ${\cal U}$ is used to ensure gauge invariance of the operator~\cite{Collins:1981uw,Collins:2002kn}.
The final state $|P_\Lambda,S_\Lambda; X\rangle$ describes the outgoing $\Lambda$ hyperon with momentum $P_\Lambda$ and spin $S_\Lambda$ together with the intermediate unobserved states.
In this paper we perform the calculation in a diquark spectator model~\cite{Nzar:1995wb,Jakob:1997wg}, which includes both the spin-$0$ (scalar diquark) and spin-$1$ (vector diquark) spectator systems~\cite{Bacchetta:2008af,Yang:2002gh}.
The quark fragmentation process (taking up quark as an example) can be modeled as $u\rightarrow \Lambda(uds) + D(\bar{d}\bar{s})$, with $D$ denoting a diquark.
The matrix element appearing in the r.h.s. of Eq.~(\ref{eq:delta1}) has the following form
\begin{align}\label{eq:diquark}
  \langle\,P_\Lambda,S_\Lambda; X|\,\bar\psi(0)|0\rangle=
  \begin{cases}
  \bar{U}(P_\Lambda,S_\Lambda)\, {\Upsilon}_s\,\displaystyle{\frac{i}{\kslash-m_q}}&\textrm{scalar diquark,} \\
 \bar{U}(P_\Lambda,S_\Lambda\,){\Upsilon}^{\mu}_v \,\displaystyle{\frac{i}{\kslash-m_q}}\,\varepsilon_{\mu}\, & \textrm{vector diquark.}
  \end{cases}
\end{align}
Here $\Upsilon_{D}$ ($D=s$ or $v$) is the hyperon-quark-diquark vertex and $\varepsilon_{\mu}$ is the polarization vector of the spin-1 vector diquark.
In our work, the vertex structure is chosen as follows~\cite{Jakob:1997wg,Yang:2017cwi}
\begin{align}
  \Upsilon_s&=\bm{1}g_s\,,\notag\\
  \Upsilon^\mu_v &={g_v\over \sqrt{3}}\gamma_5(\gamma^\mu+{P_\Lambda^\mu\over M_\Lambda}),
\end{align}
where $g_D$ ($D=s$ or $v$) is the suitable coupling for the hyperon-quark-diquark vertex.
In this work we assume that $g_s$ and $g_v$ are the same: $g_s = g_v= g_D$, and
we adopt the Gaussian form for $g_D$:
\begin{align}
g_D(k^2)={g_D^\prime \over z} e^{-k^2\over\lambda^2 z^\alpha (1-z)^\beta}
\end{align}
where $g_D^\prime$, $\lambda$, $\alpha$ and $\beta$ are the model parameters.

\begin{figure}
  \includegraphics[width=0.52\columnwidth]{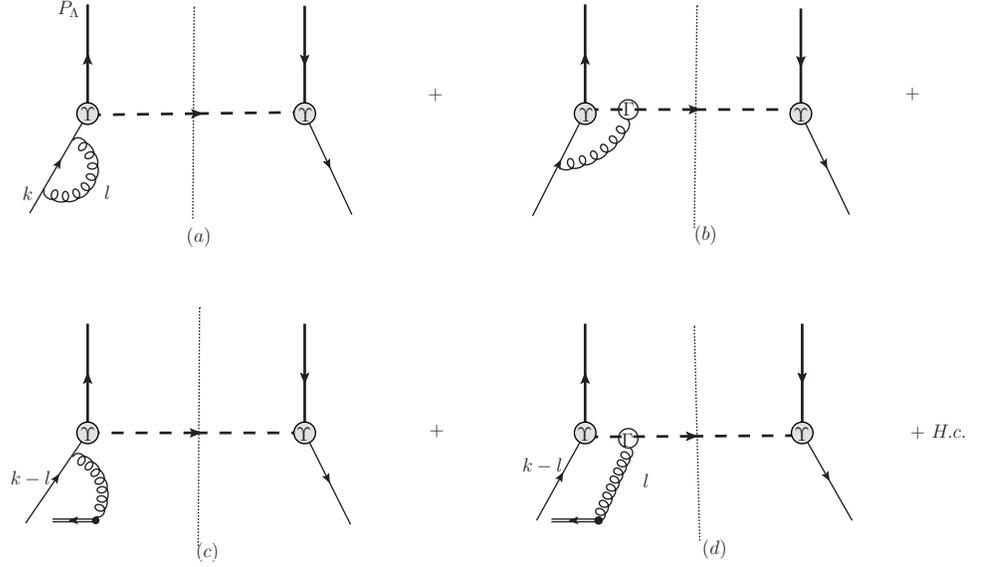}
 \caption {One loop corrections to the fragmentation of a quark to a $\Lambda$ hyperon in the spectator model. The double lines in (c) and (d) represent the eikonal lines. Here ``H.c." stands for the hermitian conjugations of these diagrams.}
 \label{loop}
\end{figure}

In the diquark model, the nonvanishing Collins function comes from the one-loop corrections which provide the necessary imaginary phases in the scattering amplitude~\cite{Brodsky:2002cx,Brodsky:2002rv}.
At one-loop level, there are four diagrams that can generate imaginary phases, as shown in Fig.~\ref{loop}.
In Figs.~\ref{loop}(b) and~\ref{loop}(d), the notation $\Gamma$ is used to depict the gluon-diquark vertex, and we apply the following rules for the vertex between the gluon and the scalar diquark ($\Gamma_s$) or the vector diquark ($\Gamma_v$):
\begin{align}
\Gamma^{\rho,a}_s &=i\,g\,T^a\,(2k-2P_\Lambda-l)^\rho\,,\\
\Gamma_v^{\rho,\mu\nu,a} &=-i\,g\,T^a\,\big{[}(2k - 2P_\Lambda -l )^\rho g^{\mu\nu}-(k-P_\Lambda-l)^\nu g^{\rho\mu}-(k-P_\Lambda)^\mu g^{\rho\nu}\big{]}\,.\label{eq:factors}
\end{align}
Here, $T^a$ is the Gell-Mann matrix, and $g$ is the coupling constant of QCD.
Since the $\Lambda$ hyperon is colorless, it is expected that the spectator diquark should have the same color as that of the parent quark.
The Feynman rules for the eikonal line as well as the vertex between the eikonal line and the gluon can be found in Refs.~\cite{Collins:1981uk,Collins:1981uw,Bacchetta:2007wc}.

Following the previous work~\cite{Yang:2017cwi} in which the fragmentation function $D_{1T}^\perp$ for the Lambda hyperon has been calculated in the same model, we perform the integration over the loop momentum $l$ with the help of the Cutkosky cutting rules.
In the l.h.s of Fig. \ref{loop}(b) and \ref{loop}(d), in principle the momentum $l$ enters the form factor for the hyperon-quark-diquark vertex with the form $g_D((k-l)^2)$.
To simplify the integration we choose that in any case the form factor $g_D$ depends only on the initial quark momentum $k$, since the main effect of the form factor is to introduce a cutoff in the high $k_T$ region. The same choice has also been used in Refs.~\cite{Bacchetta:2007wc,Lu:2015wja}.

The expression for $H^\perp_{1}$ of the $\Lambda$ hyperon, coming from the scalar diquark component, is as follows\begin{align}
H^{\perp\,(s)}_{1}(z,k_T^2) & =  {\alpha_s g_D^{\prime\,2}C_F\over (2\pi)^4 }{e^{-2k^2\over \lambda^2\,z^\alpha(1-z)^\beta}\over z^2(1-z)}{1\over (k^2-m_q^2)}\left(H^{\perp\,(s)}_{1(a)} (z,k_T^2) +H^{\perp\,(s)}_{1(b)} (z,k_T^2) +H^{\perp\,(s)}_{1(c)}(z,k_T^2)+H^{\perp\,(s)}_{1(d)} (z,k_T^2)\right)\,,
\label{eq:Collins_scalar}
\end{align}
where
\begin{align}
\begin{split}
H^{\perp\,(s)}_{1(a)} (z,k_T^2) &= {m_q\,M_\Lambda\over (k^2-m_q^2)}(3-{m_q^2\over k^2})\,I_1\,,
\end{split}\displaybreak[0]\\
\begin{split}
H^{\perp\,(s)}_{1(b)} (z,k_T^2) &= {M_\Lambda}\bigg{\{}m_q(2I_2-\mathcal{A})-M_\Lambda(\mathcal{B}-2I_2+2\mathcal{A})\bigg{\}}\,,
\end{split}\displaybreak[0]\\
\begin{split}
H^{\perp\,(s)}_{1(c)} (z,k_T^2) &=0\,,
\end{split}\displaybreak[0]\\
\begin{split}
H^{\perp\,(s)}_{1(d)} (z,k_T^2) &={M_\Lambda\over z}\bigg{\{}2(1-z)(m_q\mathcal{C}P_h^- -M_\Lambda\,\mathcal{D}P_h^-)
-z(M_\Lambda\,\mathcal{B} -m_q\mathcal{A})\bigg{\}}\,.
\end{split}\displaybreak[0]
\end{align}
Similarly, using the gluon vertex given in Eq.~(\ref{eq:factors}), we can also calculate the expression for $H^\perp_{1}$ contributed by the vector diquark component
\begin{align}
H^{\perp\,(v)}_{1}(z,k_T^2) & =  {\alpha_s g_D^{\prime\,2}C_F\over (2\pi)^4 }{e^{-2k^2\over \lambda^2\,z^\alpha(1-z)^\beta}\over z^2(1-z)}{1\over (k^2-m_q^2)}\left(H^{\perp\,(v)}_{1(a)} (z,k_T^2) +H^{\perp\,(v)}_{1(b)} (z,k_T^2) +H^{\perp\,(v)}_{1(c)}(z,k_T^2)+H^{\perp\,(v)}_{1(d)} (z,k_T^2)\right)\,,
\label{eq:Collins_vector}
\end{align}
where
\begin{align}
 \begin{split}
 H_1^{\perp\,(v)}(a) &={m_q\,M_\Lambda\over (k^2-m_q^2)}(3-{m_q^2\over k^2})\,I_1\,,
 \end{split}\displaybreak[0]\notag\\
 \begin{split}
 H_1^{\perp\,(v)}(b)&={1\over 3}\bigg{\{}2M_\Lambda[M_\Lambda(3I_2-3\mathcal{A}-\mathcal{B})+2m_qI_2]\notag\\
 &-2k\cdot P(I_2-2\mathcal{A})+{(3m_q^2-k^2)\over 4k^2}I_1+{k^2-m_q^2\over 2}(I_2-3\mathcal{A})\bigg{\}}\,,
 \end{split}\displaybreak[0]\notag\\
 \begin{split}
 H_1^{\perp\,(v)}(c) &=0\,,
 \end{split}\displaybreak[0]\notag\\
 \begin{split}
 H_1^{\perp\,(v)}(d) &={M_\Lambda\over z}[2(1-z)(m_q\mathcal{C}P_h^- -M_\Lambda\,\mathcal{D}P_h^-) -z(M_\Lambda\,\mathcal{B} -m_q\mathcal{A})]\notag\\
 &+{1\over 3M_\Lambda}\bigg{\{}4M_\Lambda(m_q\,M_\Lambda+k\cdot P)\mathcal{A}-{2M_\Lambda\over z}[(2m_qM_\Lambda+k\cdot P)\mathcal{C}P^--M_\Lambda^2\mathcal{D}P^-]\notag\\
 &+[M_\Lambda(k^2-m_q^2)\mathcal{C}P^-+2k\cdot P(m_q\mathcal{C}P^--M_\Lambda\mathcal{D}P^-)+{zm_qI_1\over 2}+{k^2-m_q^2\over 2}(m_q\mathcal{C}P^--M_\Lambda\mathcal{D}P^-)]\bigg{\}}\,.
 \end{split}\displaybreak[0]
\end{align}
Here $\mathcal{A}$, $\mathcal{B}$, $\mathcal{C}$ and $\mathcal{D}$ are functions of $k^2$, $m_q$, $m_D$ and $M_\Lambda$,
\begin{align}
\mathcal{A}&={I_1\over \lambda(M_\Lambda,m_D)} \left(2k^2 \left(k^2 - m_D^2 - M_\Lambda^2\right) {I_{2}\over \pi}+\left(k^2+M_\Lambda^2 - m_D^2\right)\right)\,,\notag\\
\mathcal{B}&=-{2k^2 \over \lambda(M_\Lambda,m_D) } I_{1}\left (1+{k^2+m_D^2-M_\Lambda^2 \over \pi} I_{2}\right),\notag\\
\mathcal{C}P_\Lambda^- &={I_{34}\over 2k_T^2} +{1\over 2zk_T^2}\left(-zk^2 + \left(2-z\right) M_\Lambda^2 + zm_D^2 \right)I_2, \notag\\
\mathcal{D}P_\Lambda^- &={-I_{34}\over 2zk_T^2} -{1\over 2zk_T^2}\left(\left(1-2z\right)k^2 + M_\Lambda^2 - m_D^2 \right)I_2. \notag
\end{align}
The functions $I_{i}$ in the above equations are defined as
\begin{align}
I_{1} &=\int d^4l \delta(l^2) \delta((k-l)^2-m_q^2) ={\pi\over 2k^2}\left(k^2-m_q^2\right)\,, \\
I_{2} &= \int d^4l { \delta(l^2) \delta((k-l)^2-m_q^2)\over (k-P_\Lambda-l)^2-m_D^2}
={\pi\over 2\sqrt{\lambda(M_\Lambda,m_D)} }  \ln\left(1-{2\sqrt{ \lambda(M_\Lambda,m_D)}\over k^2-M_\Lambda^2+m_D^2 + \sqrt{ \lambda(M_\Lambda,m_D)}}\right)\,,\\
I_{34} &= \pi\ln{\sqrt{k^2}(1-z)\over m_D}\,,
\end{align}
with $\lambda(M_\Lambda,m_D)=(k^2-(M_\Lambda+m_D)^2)(k^2-(M_\Lambda-m_D)^2)$.

In the assumption of the SU(6) spin-flavor symmetry of octet baryons, the Collins function of the $\Lambda$ hyperon for light quarks satisfies the following relations between different quark flavors and diquark types~\cite{Hwang:2016ikf,VanRoyen:1967nq,Jakob:1993th},
 \begin{align}
H_1^{\perp\textrm{u}\rightarrow \Lambda} =\,H_1^{\perp\textrm{d}\rightarrow \Lambda} ={1\over 4}H_1^{\perp(s)}+{3\over 4}H_1^{\perp(v)}\,,~~H_1^{\perp\textrm{s}\rightarrow \Lambda}=H_1^{\perp(s)}\,,
\end{align}
where u, d and s denote the up, down and strange quarks, respectively. The contributions to the Collins function $H_1^\perp$ from scalar diquark and vector diquark are given in Eqs.~(\ref{eq:Collins_scalar}) and (\ref{eq:Collins_vector}).

It is necessary to point out that the Collins function should obey the following
positivity bound~\cite{Bacchetta:2002tk,Bacchetta:1999kz}, which is a useful theoretical constraint:
\begin{align}
{|k_T|\over M_\Lambda}\left|H^{\perp}_{1}(z,\bm{k}_T^2)\right| \leq\,D_1(z,\bm{k}_T^2)\,.
\end{align}
After performing the integration over $k_T^2$, we can obtain the following approximated relation
\begin{align}
2\left|H^{\perp(1/2)}_{1}(z)\right| \leq\,D_1(z)\,, \label{eq:bound}
\end{align}
where $H^{\perp(1/2)}_{1}(z)$ is the half $k_T$-moment of the Collins function defined as
\begin{align}
H^{\perp(1/2)}_{1}(z) = z^2\,\int{d^2\bm{k}_T}{|\bm{k}_T|\over 2M_\Lambda}H^{\perp}_{1}(z,z^2\bm{k}_T^2)\,
\end{align}
and $D_1(z)=z^2\int d^2 \bm{k}_T D_1(z,z^2\bm{k}_T^2)$ is the collinear unpolarized fragmentation function.
In this work we would like to check whether the Collins function of the $\Lambda$ hyperon in our model satisfies the positivity bound, particularly, the weaker version (\ref{eq:bound}).

\begin{figure}
  \centering
  \includegraphics[width=0.48\columnwidth]{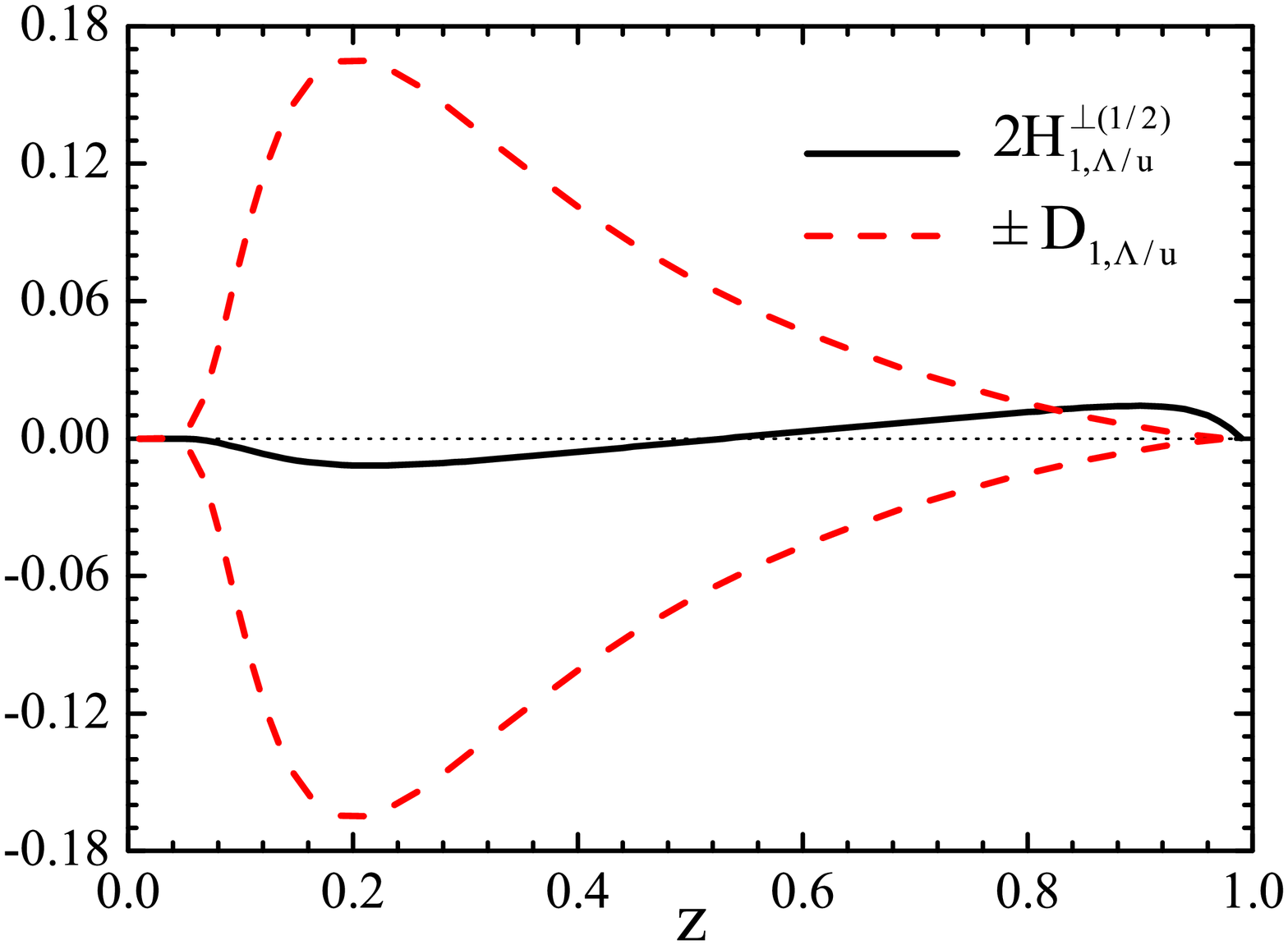}
  \includegraphics[width=0.48\columnwidth]{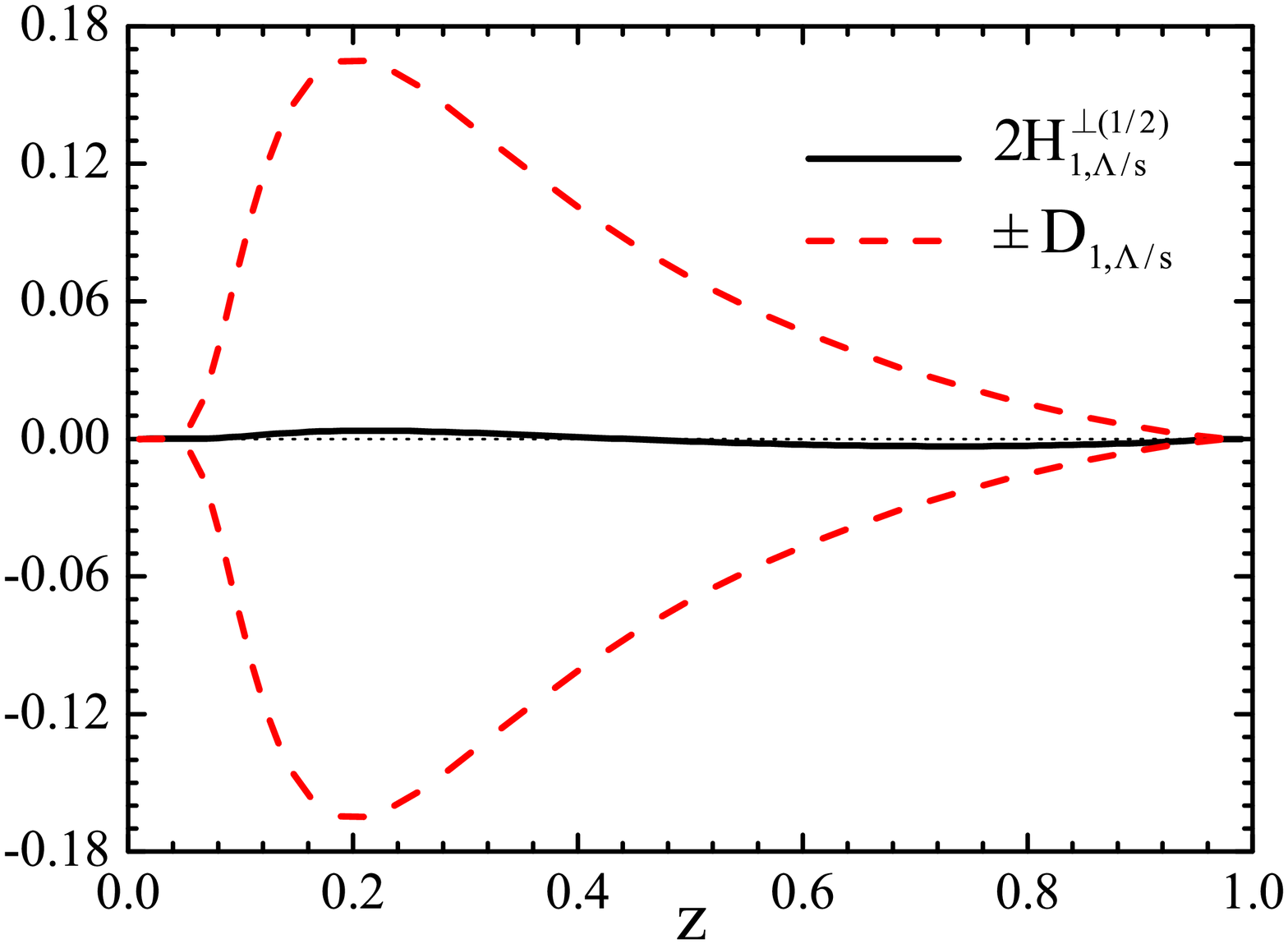}
  \caption{Left panel: the $H^{\perp(1/2)}_{1}(z)$ (multiplied by 2) (solid line) for the up quark compared with $\pm D_1(z)$ (dashed line) for the up quark at the model scale. Right panel: the $H^{\perp(1/2)}_{1}(z)$ (multiplied by 2) (solid line) and $\pm D_1(z)$ (dashed line) for the strange quark at the model scale.}
  \label{fig:compare}
\end{figure}

For the unpolarized fragmentation function $D_1(z)$ of the $\Lambda$ hyperon needed in the comparison, we apply the result from the same model in Ref.~\cite{Yang:2017cwi} as
\begin{align}
  D^{\Lambda}_1(z)=&{g_D^{\prime\, 2}\over 4(2\pi)^2}{e^{-{2m_q^2\over\Lambda^2}}\over z^4 L^2}\bigg{\{}z(1-z)((m_q+ M_\Lambda)^2 - m_D^2)\exp\biggl({-2zL^2\over (1-z)\Lambda^2}\biggr) \notag\\
  &+\bigl((1-z)\Lambda^2-2((m_q+M_\Lambda)^2-m_D^2)\bigr){z^2L^2\over\Lambda^2}\Gamma\biggl(0,{2zL^2\over(1-z)
  \Lambda^2}\biggr) \bigg{\}}\,,
\label{eq:D1}
\end{align}
To obtain this result, the mass differences among the up, down and strange quarks are neglected, and the SU(6) spin-flavor symmetry is also applied
\begin{align}
D^{u\rightarrow\Lambda}_{1}=D^{d\rightarrow\Lambda}_{1}=D^{s\rightarrow\Lambda}_{1}\equiv D_1^\Lambda\,,
\end{align}
that is, the light quarks fragment equally to $\Lambda$ for the unpolarized fragmentation function $D_1$.

In Table~\ref{tab:para}, we list the parameters~\cite{Yang:2017cwi} used in the calculations of the Collins function. The values of the parameters were obtained by fitting the model result of $D_1^\Lambda$ in the same model to the DSV parametrization for $D_1^\Lambda$~\cite{deFlorian:1997zj} at the model scale $Q_0^2 = 0.23 \mathrm{GeV^2}$. The strong coupling constant $\alpha_s$ at this scale is chosen as $0.817$.

\begin{table}
\begin{tabular}{c|c|c|c|c|c}
  \hline
  ~$m_D$~ (GeV) & $~\lambda$ (GeV)~ &~ $g_D^\prime~$ &~ $m_q$ (GeV) &~ $\alpha$ & ~$\beta$  \\
\hline\hline
  0.745 & 5.967 & 1.983 & ~0.36 (fixed)~ &~ 0.5 (fixed)~ &~ 0 (fixed)~ \\
  \hline
\end{tabular}
\caption{Values of the parameters used in the spectator diquark model~\cite{Yang:2017cwi}.
The values of the last three parameters are fixed.}
\label{tab:para}
\end{table}

In Fig.~\ref{fig:compare}, we plot the numerical result of $H^{\perp(1/2)}_{1}(z)$ (multiplied by a factor of 2) of the $\Lambda$ hyperon (solind lines), compared with the unpolarized $\Lambda$ fragmentation function $\pm D_1^\Lambda(z)$~(dashed lines) in the same model.
The left panel shows the result for the up/down quark, while the right panel depicts the result for the strange quark.
From the curves, one can find that the size of $H^{\perp(1/2)}_{1}(z)$ for the up and down quarks is around several percent.
Particularly, the sign of $H^{\perp(1/2)}_{1,\Lambda/\textrm{u}}(z)$ is negative in the small $z$ region~($0<z<0.5$), while it turns to be positive in the large $z$ region~($0.5<z<1$).
That is, there is a node in the $z$-dependence of the Lambda Collins function for the up and down quarks. This is different from the Collins function of the pion for which no node appears.
We also find that $H^{\perp(1/2)}_{1}(z)$ for the strange quark is consistent with zero.
Finally, our model result of $H^{\perp}_{1}$ for the up and down quarks does not always satisfy the positivity bound, i.e., in the large $z$ region ($z>0.82$) the positivity bound is violated.
We note that similar violations of the positivity bound were also observed in Refs.~\cite{Pasquini:2014ppa,Wang:2017onm,Yang:2017cwi}.
An explanation was given in Ref.~\cite{Pasquini:2011tk}, stating that the violation may arise from the fact that T-odd TMD distributions or fragmentation functions are evaluated to $\mathcal{O}(\alpha_s)$, while in model calculations T-even TMD functions are usually truncated at the lowest order.

Since the energy scale in experiments is much higher than the model scale, it is important to include the QCD evolution of fragmentation functions to obtain reliable results for physical observables.
In Refs.~\cite{Yuan:2009dw,Kang:2010xv}, the evolution equation for the twist-3 fragmentation function $\hat{H}(z)$ has been studied. This fragmentation function is proportional to the first $k_T$-moment of Collins function via the relation
\begin{align}
\hat H(z) =z^2 \int d^2 \bm k_T {\bm k_T^2\over M_\Lambda}  H_1^\perp(z,\bm k_T^2) = 2M_\Lambda H_1^{\perp (1)}(z)
\end{align}
The evolution kernel for $\hat H(z)$ has a rather complicated form. Following Ref.~\cite{Kang:2015msa}, in this work we only consider the homogenous terms~\cite{Kang:2010xv} in the kernel, which have the same form of the evolution kernel for the transversity distribution function $h_1$:
\begin{align}
P^{h_1}_{qq}=C_F\left(\frac{2z}{(1-z)_+}+\frac{3}{2}\delta(1-z)\right). \label{eq:kernel}
\end{align}

\begin{figure}
  \centering
  \includegraphics[width=0.48\columnwidth]{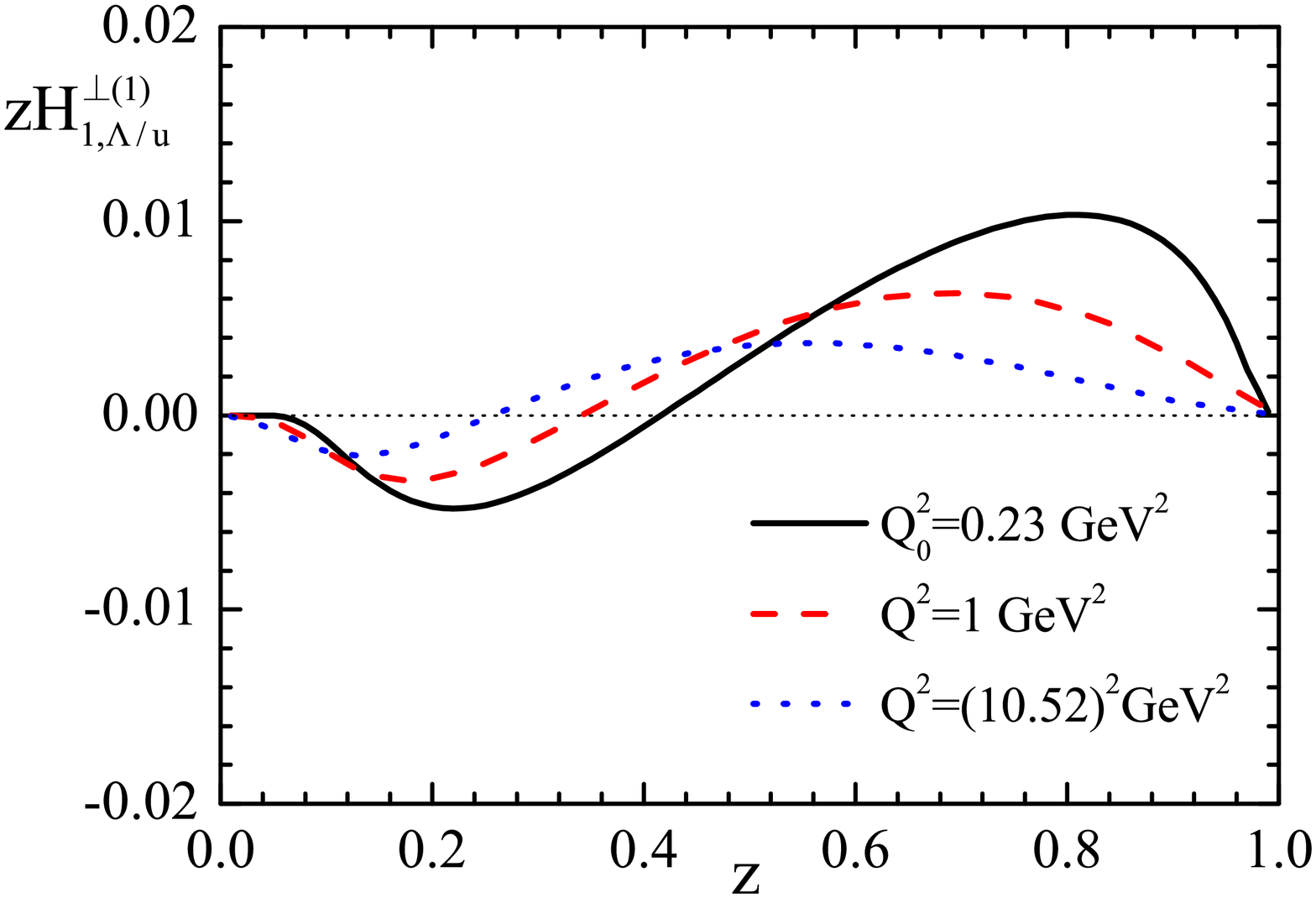}
  \includegraphics[width=0.48\columnwidth]{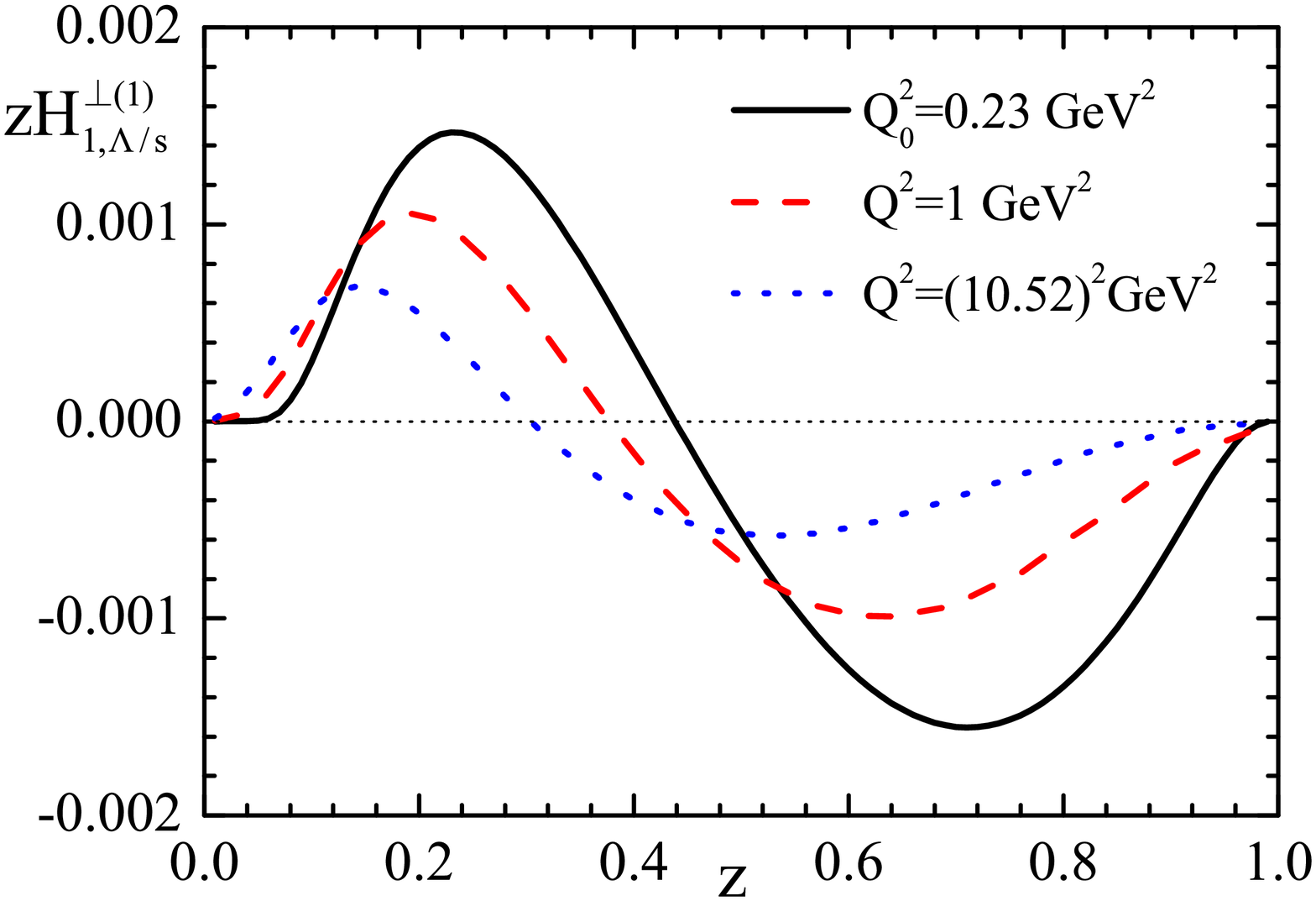}
    \caption{The Collins function of the $\Lambda$ hyperon at three different energy scales: $Q_0^2=0.23\mathrm{GeV}^2$~(solid lines), $Q^2=1\mathrm{GeV}^2$~(dashed lines) and $Q^2=(10.52)^2\mathrm{GeV}^2$~(dotted lines). Left panel: $zH^{\perp(1)}_{1}(z)$ for the up and down quark; right panel: $zH^{\perp(1)}_{1}(z)$ of the strange quark.}
    \label{fig:evolution1}
\end{figure}

\begin{figure}
  \centering
  \includegraphics[width=0.48\columnwidth]{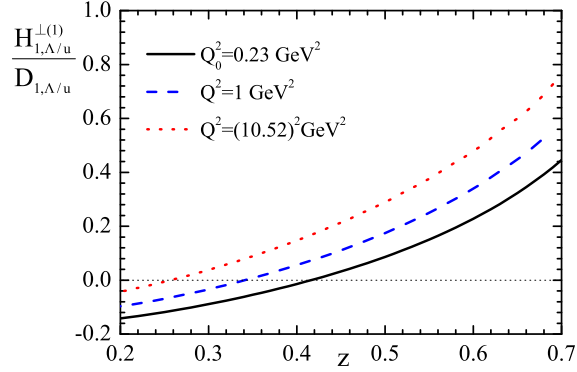}
  \caption{The ratio $H^{\perp(1)}_{1,\Lambda/\textrm{u}}(z,Q^2)/D_1^{\Lambda}(z,Q^2)$ at three different energy scales: $Q_0^2=0.23\mathrm{GeV}^2$~(solid lines), $Q^2=1\mathrm{GeV}^2$~(dashed lines) and $Q^2=(10.52)^2\mathrm{GeV}^2$~(dotted lines).}
  \label{fig:evolution2}
\end{figure}

We apply the evolution package {\sc{QCDNUM}}~\cite{Botje:2010ay} and custom the code to include the kernel in Eq.~(\ref{eq:kernel}) to perform the evolution of $H^{\perp(1)}_{1}(z)$.
In Fig.~\ref{fig:evolution1}, we plot the first $k_T$-moment of the lambda Collins function $H^{\perp(1)}_{1}(z)$, which plays the role in the weighted azimuthal asymmetry.
The left and right panels show the results for the up/down quark and the strange quark at three different energy scales. The solid lines depict the model results at the initial scale $Q_0^2=0.23\mathrm{GeV}^2$, while the dashed and dotted lines show the results at $Q^2=1\mathrm{GeV}^2$ and $Q^2=10.52^2\mathrm{GeV}^2$ after applying the evolution equation for $H_1^{\perp(1)}(z)$.
From the curves, we can see that the evolution effect for $H^{\perp(1)}_{1}(z)$ is significant, i.e., the evolution changes the shape and the size of the fragmentation functions at different $Q$ values.
It drives the peaks of $H^{\perp(1)}_{1}(z)$ to the lower $z$ region with increasing $Q$.
At higher scale, the node of $H^{\perp(1)}_{1}(z)$ for the up or down quark also moves to the lower $z$ region.
The similar tendency also appeared in the transversity distribution function of the nucleon for the up quark in Ref.~\cite{Bacchetta:2008af}.
In order to demonstrate the evolution effects of fragmentation functions in the weighted azimuthal asymmetries, in Fig.~\ref{fig:evolution2} we also plot the ratio $H^{\perp(1)}_{1}(z,Q^2)/D_1^\Lambda(z,Q^2)$ for up quark at three scales.
We find that, in the region $0.2<z<0.7$, the ratio $H^{\perp(1)}_{1,\Lambda/\textrm{u}}(z,Q^2)/D_1^{\Lambda}(z,Q^2)$  increases with the increasing $z$ at any energy scale.

\section{$q_T^2$-Weighted $\cos 2 \phi_0$ asymmetry in the $e^+ e^-\to \Lambda \bar\Lambda X$ process}
\label{Sec:weighted}

Using the Collins function calculated in Sec.~\ref{sec:model} and the unpolarized fragmentation function $D_1$ in Eq.~(\ref{eq:D1}),
in this section, we will numerically estimate the $q_T^2$-weighted $\cos 2\phi_0$ azimuthal asymmetry in the process
\begin{align}
\label{eq:process}
e^+ + e^-\longrightarrow \Lambda +\overline{\Lambda} + X,
\end{align}
where the two leptons $e^+$~(with momentum $l$) and $e^-$~(with momentum $l^\prime$) annihilate into a photon with momentum $q=(l+l^\prime)$.
The photon then produces a quark-antiquark pair, which fragments into the final state hadron pair $\Lambda,\ \bar{\Lambda}$ and other unobserved states.
In the unpolarized process, the double Collins effect shows up at the leading order (in $1/Q$ expansion), with a $\cos 2\phi_0$ modulation in the differential cross section~\cite{Boer:2008fr,Boer:1997mf}:
\begin{align}
\frac{d\sigma (e^+ e^- \rightarrow \Lambda\bar{\Lambda} X)}{dz_1dz_2d\Omega {d}^{2}{\bm{q}}_{T}}=&\frac{3\alpha_{em}^{2}}{Q^2}z_1^2z_2^2\bigg{\{ }A(y)\mathcal{F}[D_1\overline{D}_1]+\nonumber
\\&B(y)\cos2\phi_0\mathcal{F}\left[(2\bm{\hat{h}}\cdot\bm{k}_{T} \bm{\hat{h}}\cdot \bm{p}_{T}-\bm{k}_{T}\cdot \bm{p}_{T})\frac{{H}^{\perp}_1 \overline{H}^\perp _1}{{M}_{\Lambda}{M}_{\bar{\Lambda}}}\right]\bigg{\}}, \label{eq:conv}
\end{align}
where $\phi_0$ is the azimuthal angle of the $\Lambda$ hyperon in the c.m frame of the incoming $e^+ e^-$ pair, with the $z$ axis along the momentum of $\bar{\Lambda}$~\cite{Boer:1997mf}.
In Eq.~(\ref{eq:conv}) we adopt the notation
\begin{align}
\mathcal{F}[\omega D\bar{D}]=&\sum_a e_a^2\int d^2\bm{k}_{T}{d}^{2}\bm{p}_T\delta
^2(\bm{k}_T+\bm{p}_T-\bm{q}_T)\nonumber\omega D^a(z_1,z_1^2\bm{k}_T^2)\bar{D}^{\bar{a}}(z_2,z_2^2\bm{p}_T^2)
\end{align}
to express the convolution of transverse momenta with $\omega$ being an arbitrary function.
$z_1$ and $z_2$ are the longitudinal momentum fraction of the produced $\Lambda$ and $\bar{\Lambda}$ from the parent quarks; $\bm{q}_T$, $\bm{k}_T$ and $\bm{p}_T$ are the transverse momenta of the photon, the fragmenting quark and antiquark, respectively. The unit vector $\bm{\hat{h}}$ is defined as $\bm{\hat{h}}=\frac{\bm{q}_{T}}{|\bm{q}_{T}|} = \frac{\bm{q}_{T}}{q_T}$. $\bar{D}_1$, $\bar{H}^\perp _1$ denote the corresponding fragmentation functions of the antiquark $\bar{q}$ to the $\bar{\Lambda}$ hyperon.
The kinematical factors $A(y)$ and $B(y)$ have the following form~\cite{Boer:2008fr}
\begin{align}
&A(y)=\left(\frac{1}{2}-y+y^2\right) \ \stackrel{\textrm{c.m.}}{=}
\ \frac{1}{4}\left(1+\cos^2 \theta\right),\nonumber\\
&B(y)  \ =
\ y(1-y)  \ \stackrel{\textrm{c.m.}}{=}
\ \frac{1}{4} \sin^2 \theta, \nonumber
\end{align}
where $\theta$ is the angle of between the momentum of the incoming lepton $\bm{l}$ and the $z$-axis.

\begin{figure}
  \centering
  \includegraphics[width=0.49\columnwidth]{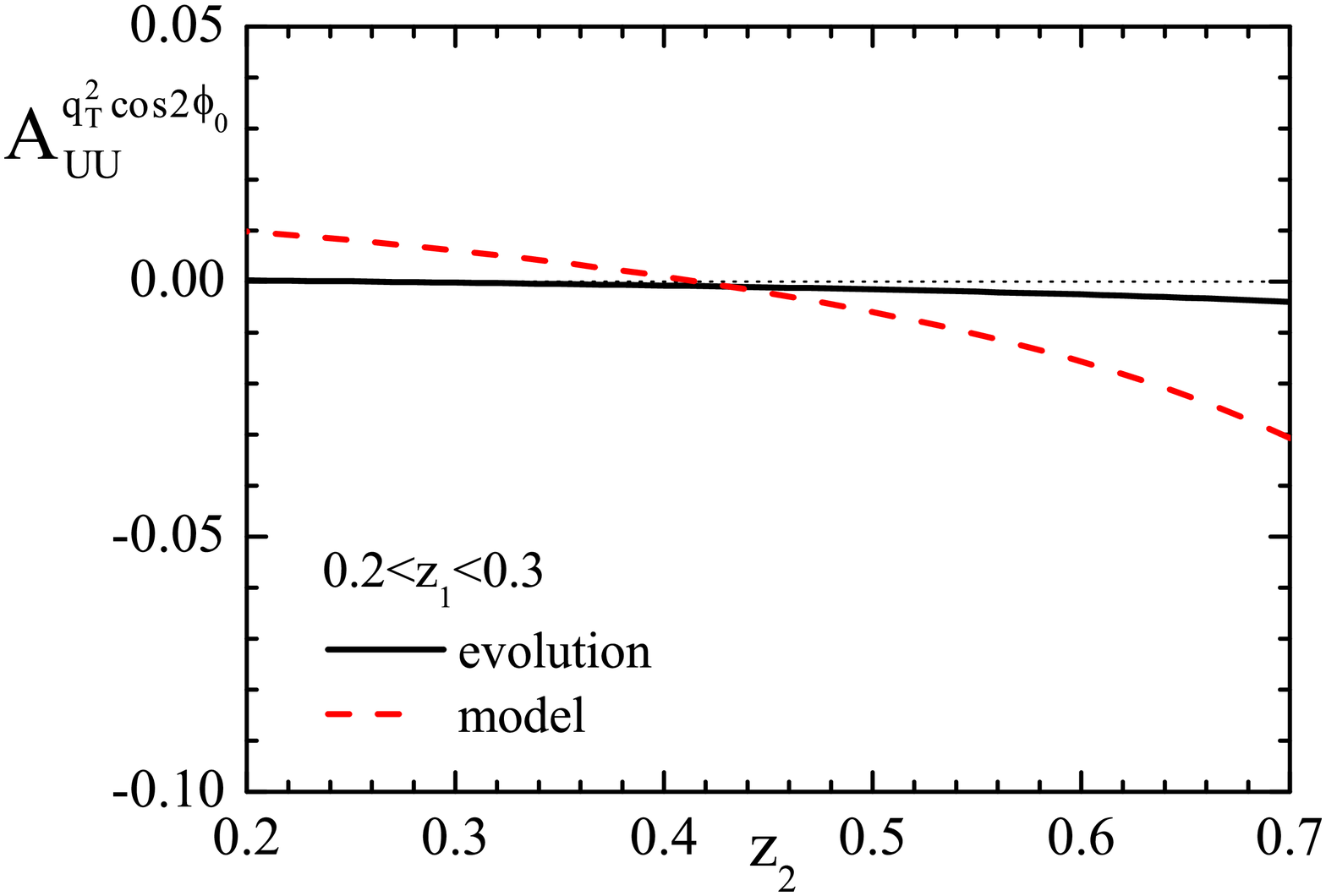}
  \includegraphics[width=0.49\columnwidth]{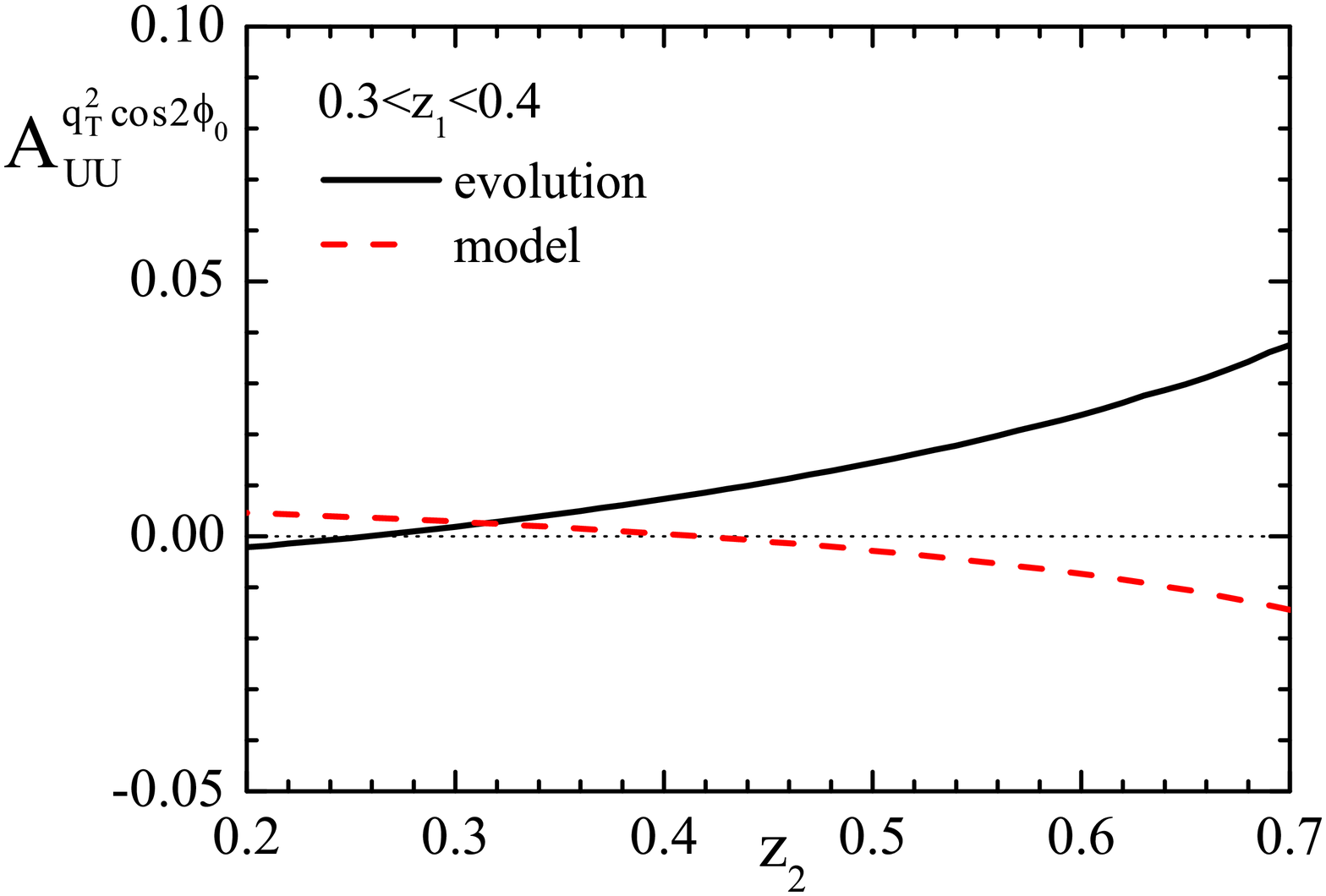}\\
  \includegraphics[width=0.49\columnwidth]{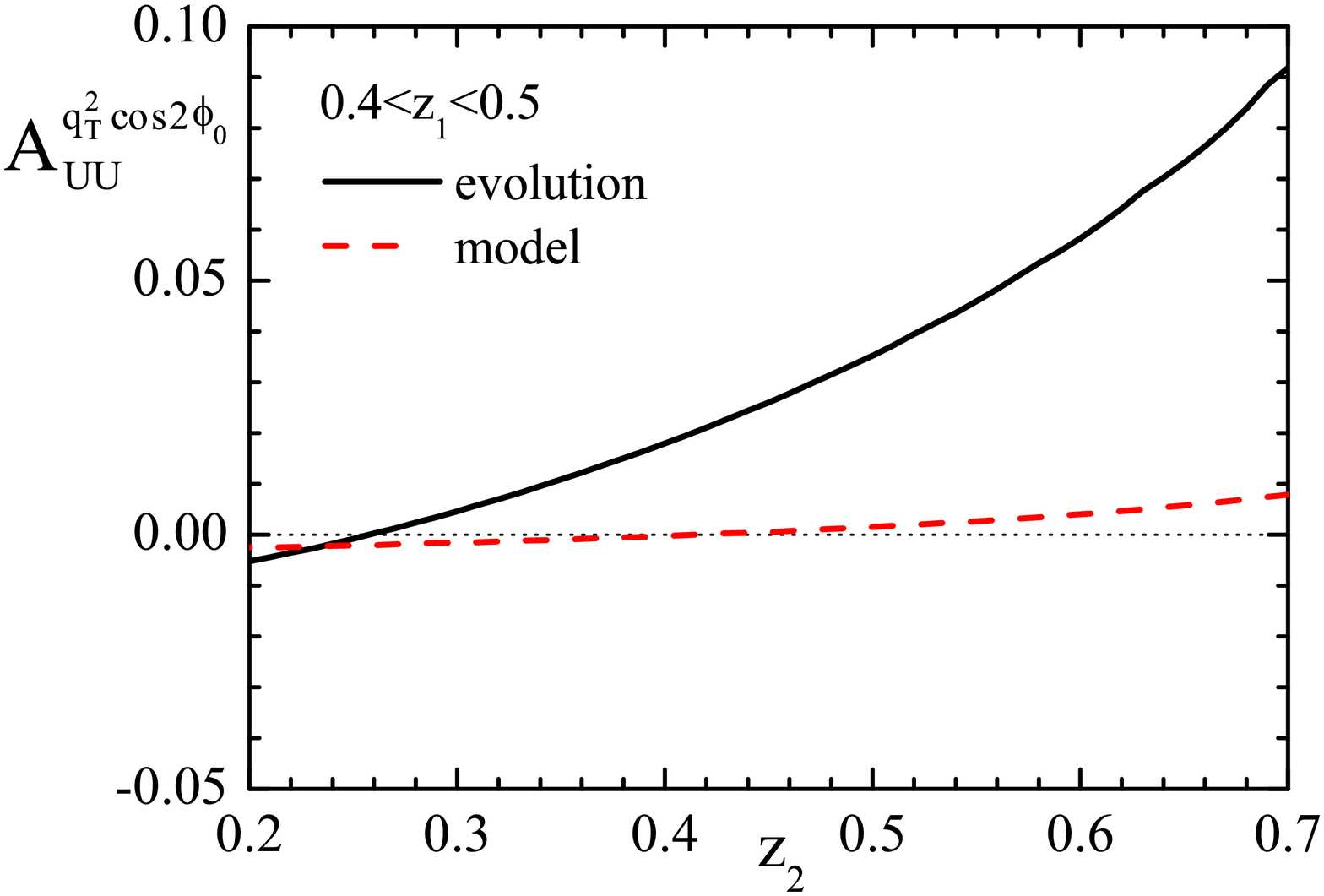}
  \includegraphics[width=0.49\columnwidth]{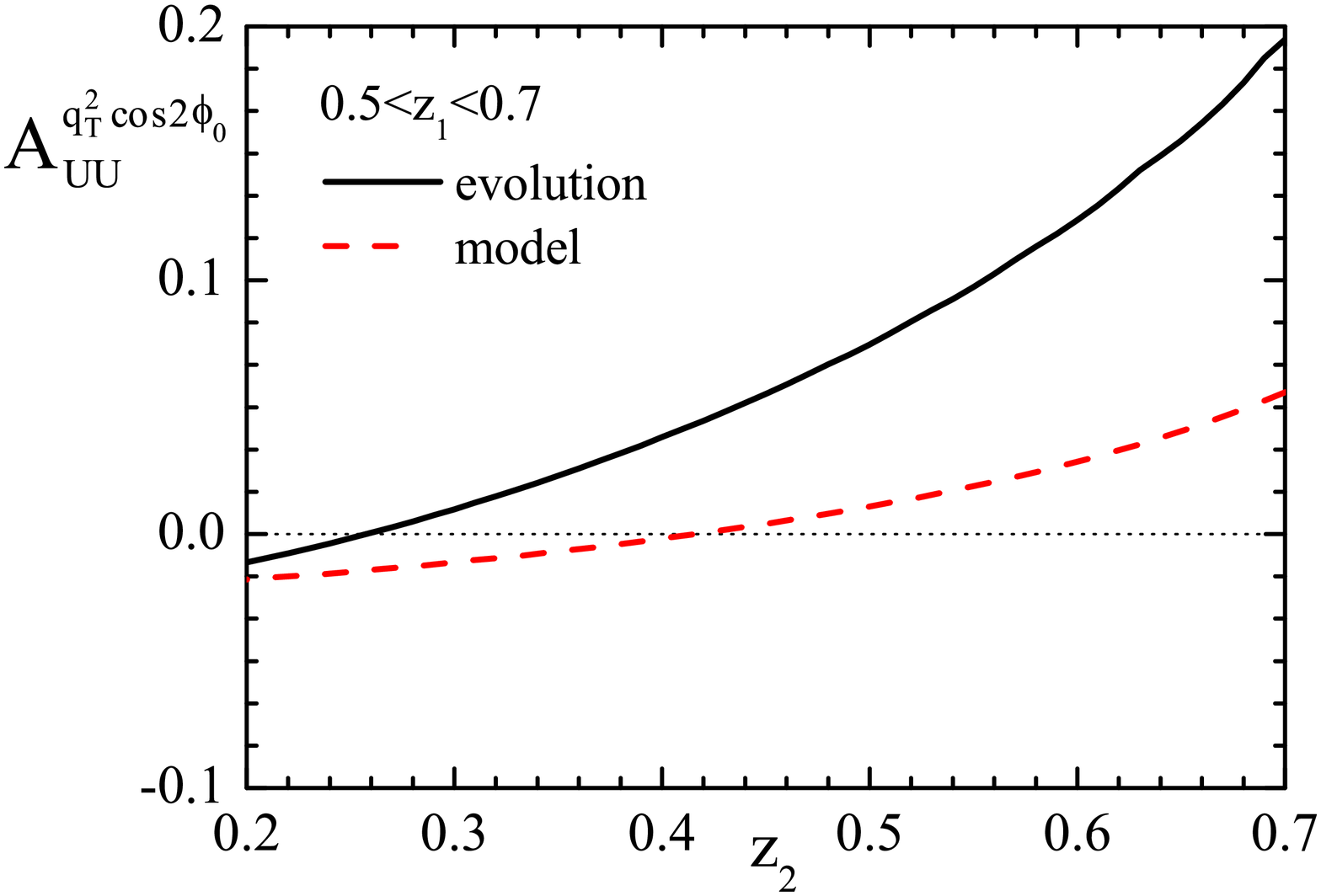}\\
  \caption{The weighted Collins asymmetry for the $e^+ e^-$ annihilating to $\Lambda\bar{\Lambda}$ process as the function of $z_2$ with $z_1$ integrated in each bin. Dashed lines represent the asymmetry assuming the fragmentation function do not evolve with energy scales. Solid lines denote the asymmetry considering both the evolution effect of $D_1$ and $H_1^{\perp(1)}$.}
  \label{fig:asy}
\end{figure}

Since the differential cross section in Eq.~(\ref{eq:conv}) contains the convolutions of the TMD fragmentation functions depending on $z$ and the transverse momentum, one needs to study the transverse momentum dependence behavior of the fragmentation functions, which is described by TMD evolution and is complicated to analyze. As an alternative approach, the transverse-momentum weighted asymmetries, for which the $k_T$-moments of the fragmentation functions play an important role, have been proposed in Refs.~\cite{Kotzinian:1995cz,Boer:1997mf}.
The weighted cross section in the process $e^+ +  e^- \to \Lambda + \bar{\Lambda} + X$ can be defined as~\cite{Boer:2008fr}
\begin{align}
\langle W \rangle=\int d^2\bm{q}_T W \frac{d\sigma(e^+ e^- \rightarrow \Lambda\bar{\Lambda} X)}{dz_1dz_2d\Omega {d}^{2}{\bm{q}}_{T}},
\end{align}
where $W$ is the weighting function.

Choosing the proper weighting function ${q_T^2\cos 2\phi_0 /4M^2_\Lambda}$, one can define the weighted Collins asymmetry as
\begin{align}
A^{q^2_T\cos 2\phi_0}_{UU}&=\frac{\langle\frac{q_T^2}{4M^2_\Lambda}\cos 2\phi_0\rangle}{\langle 1 \rangle}\nonumber\\
&=\frac{B(y)}{A(y)}\frac{\sum_a e_a^2 H_1^{\perp(1)q}(z_1)\bar{H}_1^{\perp(1) \bar{a}}(z_2)}{\sum_a e_a^2 D_1^a(z_1)\bar{D}_1^{\bar{a}}(z_2)}.
\end{align}
We apply the above expression to estimate the $\cos2\phi_0$ azimuthal asymmetry in the process $e^+ e^- \rightarrow \Lambda\bar{\Lambda}X$ at $Q=10.52\,\mathrm{GeV}$, which is the scale of the Belle measurement~\cite{Seidl:2008xc} and which is also close to the kinematics of the BaBar measurement.
As the energy scales in these experiments are much higher than the model scale, we need to take into account the QCD evolution effects of the fragmentation functions.

To study the impact of the evolution effect, we adopt two different ways to calculate the weighted asymmetry $A^{q^2_T\cos 2\phi_0}_{UU}$. One is to assume that all the fragmentation functions do not evolve with energy scale, which is an extreme condition. The other is to apply the evolution kernel in (\ref{eq:kernel}) for $H_1^{\perp(1)}(z)$ and the DGLAP evolution for $D_1(z)$.
For the factor of ${B(y)/A(y)}=\frac{\sin^2\theta}{1+\cos^2\theta}$ at Belle, the mean value in each $(z_1,z_2)$ bin is given in Ref.~\cite{Seidl:2008xc}, here we take 0.7 as a rough estimate.

In Fig.~\ref{fig:asy}, we plot the weighted azimuthal asymmetry $A^{\bm{q}_T^2\cos2\phi_0}_{UU}$ contributed by the double Collins effect as the function of $z_2$ for four $z_1$ bins: $[0.2, 0.3]$, $[0.3, 0.4]$, $[0.4, 0.5]$ and $[0.5, 0.7]$.
In each bin the variable $z_1$ is integrated.
Note that the bins for $z_1$ in this work are slightly different from those in Ref.~\cite{Seidl:2008xc}.
Since the Collins function in our model violates the positivity bound at large $z$ region~($z>0.8$), we avoid the bin $[0.7,1]$.
The dashed lines in Fig.~\ref{fig:asy} show the asymmetry under the extreme assumption in which the evolution of the fragmentation functions are ignored, while the solid lines denote the asymmetry in the case the evolution effects of both $D_1$(z) and $H^{\perp(1)}_{1}(z)$ are included.
The solid lines indicate that the weighted asymmetry is positive, and it is sizable in the large $z_1$ or $z_2$ region.
In addition, the asymmetry increases with increasing $z$.
Similar results were also find in the case of pion pair production in $e^+e^-$ annihilation~\cite{Abe:2005zx,Bacchetta:2007wc}.
Comparing the solid lines and dashed lines, we can also see that the evolution effects significantly affect the weighted $\cos2\phi_0$ asymmetry in the $e^+ e^-\rightarrow\Lambda\bar{\Lambda}X$ process thereby it should not be neglected.

\section{Conclusion}
\label{sec:conclusion}

In this work, we investigated the T-odd Collins function $H_1^{\perp}$ of the $\Lambda$ hyperon for light quarks as well as its contribution to the $q_T^2$-weighted $\cos 2\phi_0$ azimuthal asymmetry in $e^+e^-\rightarrow \Lambda\bar{\Lambda}X$ process.
We calculated the Collins function of the $\Lambda$ hyperon in the diquark spectator model by considering both the scalar and vector diquark components.
In the calculation we adopted a Gaussian form factor for the hyperon-quark-diquark vertex, and we apply the values of the model parameters fitted from the DSV parametrization at the initial scale $Q_0^2=0.23\mathrm{GeV}^2$.
The numerical result shows that the lambda Collins function for the up and down quark dominates over that for the strange quark.
We also calculated the QCD evolution of the first $k_T$-moment of the lambda Collins function and found that the evolution effects significantly alter $H_1^{\perp(1)}(z)$.
By applying the model results for $H_1^{\perp(1)}(z)$ and $D_1(z)$, we estimated the $q_T^2$-weighted $\cos2\phi_0$ azimuthal asymmetry contributed by the double Collins effect in the unpolarized $e^+e^-\rightarrow\Lambda\bar{\Lambda}X$ process at $Q=10.52$ GeV in two scenarios: one is to take into account evolution of both $H_1^{\perp(1)}(z)$ and $D_1(z)$; the other is to neglect any scale dependence of fragmentation functions.
We found that in the former case, the asymmetry is positive and increases with increasing $z_1$ and $z_2$, which is similar to the case of the charged pion pair production in $e^+ e^-$ annihilation.
Therefore it is feasible to measure this asymmetry through the Belle and BaBar experiments.
We also found that the evolution effects significantly change the shape and size of the asymmetry.
Our study may provide useful information on the lambda fragmentation function as well as the nonperturbative origin of the azimuthal asymmetry in $e^+ e^-$ annihilation.

\section{Acknowledgements}

This work is partially supported by the NSFC (China) grant 11575043, by the Fundamental Research Funds for the Central Universities of China. Y. Y is supported by  the Scientific Research Foundation of Graduate School of Southeast University (Grant No. YBJJ1770) and by the Postgraduate Research \& Practice Innovation Program of Jiangsu Province (Grants No. KYCX17\_0043). X. W is supported by the Scientific Research Foundation of Graduate School of Southeast University (Grants No.~YBJJ1667). X. W and Y. Y contributed equally to this work and should be considered as co-first authors.

\end{document}